\documentstyle[12pt]{article}
\def\lapproxeq{\lower .7ex\hbox{$\;\stackrel{\textstyle <}{\sim}\;$}}
\def\gapproxeq{\lower .7ex\hbox{$\;\stackrel{\textstyle >}{\sim}\;$}}

\begin{document}

\titlepage

\begin{flushright} RAL-94-071 \\ June 1994
\end{flushright}

\begin{center}
\vspace*{2cm}
{\large{\bf  The Spin Dependence of Diffractive Processes and Implications
for the Small $x$ Behaviour of $g_1$ and the Spin Content of the Nucleon}}

\end{center}

\vspace*{.75cm}
\begin{center}
F.E.\ Close and R.G.\ Roberts \\
Rutherford Appleton Laboratory, \\ Chilton, Didcot OX11 0QX, England. \\
\end{center}

\vspace*{1.5cm}

\begin{abstract}
We show that if the Lorentz transformation properties of diffraction are
other than scalar, the $x \rightarrow 0$ behaviour of $g_1(x,Q^2)$ can
grow. We compare with new data on $g_1^p$ from SMC,
assess implications for
sum rules and for future studies of sea polarisation.
\end{abstract}

\newpage

\noindent{\large{\bf Introduction}}
\vskip 0.5cm
The measurement of the net quark spin content of the proton and neutron
by deep inelastic polarised leptoproduction requires an integral over the
structure function $g_1(x,Q^2)$. This includes an extrapolation to high
energies, or equivalently $x=0$, which has tended to be based on Regge theory
and the assumed dominance of an $a_1$ trajectory. In this case

\begin{equation}
g_1 \approx x^{-\alpha_{a_1}}, \quad  x\rightarrow 0
\label{g1approx}
\end{equation}
where $\alpha_{a_1}$ is the intercept of the $a_1$ Regge trajectory.
This has been assumed to lie in the range $-0.5< \alpha_{a_1} <0$
and errors on the extrapolation have incorporated this range of
values for the intercept.

The current value inferred for the net spin, based on all measurements
with a proton target \cite{emc2,smc,e130} is
\begin{equation}
\Delta q =0.30 \pm 0.07(stat) \pm 0.10(syst)
\label{delsigsmc}
\end{equation}
This is consistent with  the historically measured values though the central
value has increased significantly from the original \cite{emc1} estimate
of a value consistent with zero.

A significant part of the increase in the inferred value (which is
today some two standard deviations below the naive quark model expectation
in the absence of strange quark and/or gluon polarisation) is due to
the increase in the magnitude of the measured or inferred
data on $g_1(x)$ at small $x$. An important ingredient in this is the
fact that the $g_1(x)$ is constructed from a measured polarisation asymmetry
which has to be multiplied by the unpolarised structure function, $F_2$,
\begin{equation}
g_1(x,Q^2) = A_1(x,Q^2) F_2(x,Q^2) / 2x(1+R(x,Q^2))
\label{g1defn}
\end{equation}
and $F_2$
is now known to grow in magnitude at small $x$ \cite{nmc,hera}
as well as being intrinsically
larger in overall normalisation than believed originally \cite{fcsing}.

A superficial glance at the SMC \cite{smc} data hints that $g_1^p(x)$
may be rising
for $x<0.01$ (which is a result of $A(x)$ being roughly constant while the
unpolarised structure function is growing). If this trend is confirmed,
and if it continues to smaller values of $x$, then the naive Regge pole
extrapolation will be inadequate.

This leads us to the main point of this paper: {\it what empirical knowledge
or theoretical constraints are there on the high energy behaviour (or
small $x$ behaviour) of spin dependent total cross sections (polarised
structure
functions)?} It seems to us that the literature allows the
possibility of considerable polarisation dependence in the diffractive
region out to large energies and small values of $x$. We shall consider four
examples: an empirical study by Martin \cite{am}, a generalisation
 of Froissart's heuristic derivation of high energy dependence to spin
dependence \cite{froiss}, the $x \rightarrow 0$ behaviour of $g_1(x)$ in
the double log approximation (DLA) of QCD,
and a specific model of the Pomeron following from ideas of Donnachie and
Landshoff \cite{donlan}.
We then compare these and other models with the data at
the smallest $x$ values and evaluate their consequences for the sum rule. We
close by assessing what future possibilities there are of improving on
the empirical evaluation of the polarisation at small $x$.
\vskip 1.0cm
\noindent{\large{\bf Limits from proton-proton scattering}}
\vskip 0.5cm
First, it is worth noting that the measurement of $g_1$ is unique in that
it is the only measurement of a high energy spin dependent total cross section
in hadron physics. Martin \cite{am} has shown that one can place a limit
on the polarisation dependence for high energy $p p$ scattering since
 the p-p total cross sections are measured in colliders by combining two of
the three quantities

1) the luminosity ${\cal L}$

2) the total number of events per second ${\cal L}\sigma(total)$

3) the extrapolated number of elastic events per second at $t=0$ i.e.
${\cal L}\frac{d\sigma}{dt}|_{t=0}$.
If spin effects are unimportant this is related
to ${\cal L}\sigma_{tot}^2$ once the real part is known; conversely
a difference between these arises if spin effects are large.

When these comparisons are applied to ISR data one finds \cite{am}
 that the ratio
$\sigma^{\uparrow \uparrow}/\sigma^{\uparrow \downarrow}$
could lie anywhere between
3/4 and 4/3. At the $Sp\bar{p}S$ the constraints are much poorer
($1 \over 2$ to 2 in ratio). Thus one may conclude that spin asymmetries
$A=\Delta \sigma/\sigma$
could be as large as 0.14 at ISR energies or 0.33 at the  $Sp\bar{p}S$.
These data offer no reason to require a small asymmetry in either
polarised $p p$ or (virtual) photoproduction and highlight the importance of
these latter as pioneering measures of high energy spin dependence.
They also encourage interest in possible proton polarisation at RHIC and
measurment of the energy dependence of the asymmetry.
\vskip 1.0cm
\noindent{\large{\bf Asymptotic bounds and log $x$ dependence}}
\vskip 0.5cm
Theoretical bounds exist for the rise with energy of total cross
sections (unpolarised), namely that \cite{froiss,frm}
\begin{equation}
\sigma \leq \log^2 s
\label{log2}
\end{equation}
Froissart showed how this bound is realised in an heuristic model.
Consider two particles scattering via a potential parametrised
as
\begin{equation}
V(r) =  gs^N \exp(-\mu r)
\label{pot}
\end{equation}
where $N$ is near to unity (as in simple diffractive Pomeron exchange)
and $\mu$ is an inverse length, or mass, scale. Clearly the effective
range will grow as $s$ increases. The scaling behaviour of the effective
range, $R$, with energy follows by setting $V(R)=1$ and hence
\begin{equation}
R \approx (\log g + N \log s )/\mu
\label{rapprox}
\end{equation}
in which case the cross section reaches the Froissart bound
\begin{equation}
\sigma =\pi R^2 \approx \log ^2 s
\label{froissart}
\end{equation}

The spin dependence of the cross section depends on the Lorentz nature
of the potential. Only for the case of a {\it scalar} is there
no spin dependence in the diffractive scattering; in
this case all spin dependence would follow from the (non-diffractive)
processes such as $a_1$ Regge pole exchange as in
the present assumed pole parametrisations \cite{smc,ek}.

An alternative picture, which may be rooted in ideas from QCD where
diffractive scattering is driven by multi gluon exchange, will in
general have non trivial Lorentz structure, in particular {\it vector}
exchange. (A particular model of diffractive scattering due to
Donnachie and Landshoff \cite{donlan} makes an analogy between
Pomeron and photon
such that the Pomeron is assumed to couple
via a vector $\gamma_{\mu}$ \cite{ln}).

The effective potential has a non leading spin dependence \cite{co}
\cite{gromes}
\begin{equation}
V(r) + \vec{\sigma} \cdot \vec{\sigma}  \nabla^2V(r) /s
\label{effpot}
\end{equation}
which is reminiscent of the hyperfine low energy interaction
in atomic hydrogen. If one now includes this in the potential
argument above
\begin{equation}
V \approx g(s^N \pm \mu^2s^{N-1}) exp(-\mu r)
\label{vextra}
\end{equation}
and so for large $s$ one finds that eq(\ref{rapprox}) generalises to
\begin{equation}
R^2 \approx N^2\log ^2 s \pm \frac{2N \mu^2 \log s}{s }
\label{r2approx}
\end{equation}
implying that the spin asymmetry can behave as
\begin{equation}
A \approx \frac{1}{s \log s}
\label{as}
\end{equation}
or equivalently that
\begin{equation}
\Delta \sigma \approx \frac{\log s}{s}
\label{delsig}
\end{equation}
If one is allowed to identify $s$ with $1/x$ then these imply that $g_1$
is limited by
\begin{equation}
g_1(x \rightarrow 0) \approx -\log x
\label{g1logx}
\end{equation}

Of course there is no reason to expect the Froissart bound to be saturated
but since the new small $x$ data on both $F_2$ and $g_1$
are interestingly large we need
to examine what limits can be set on the behaviour of $g_1$ in this region.
An explicit calculation of the
spin dependent diffractive scattering in the Landshoff Donnachie
model (which does not saturate the Froissart bound in the unpolarised case)
does manifest the $\log x$ behaviour, even at the presently attainable
values of $x$
viz \cite{basslan}
\begin{equation}
g_1(x) \sim (1 + 2 \log x)
\label{g1bl}
\end{equation}

It is interesting to consider what would occur if the potential transformed as
an {\it axial vector}. In this case there is spin dependence in leading order
\cite{gromes} and the scattering is attractive only in one spin state
(parallel or antiparallel depending on the overall sign). In this case
the limiting behaviour is extreme
\begin{equation}
xg_1 \sim \log ^2 x
\label{g1log2}
\end{equation}
in which eventuality
the integral (spin sum) diverges. Physically this would imply that
the sea is produced in one polarisation state only. This may appear
artificial and lies outside known QCD mechanisms; we shall
not pursue this possibility further even though it is allowed a priori.

In general we note that if
the elastic scattering potential
transforms other than as a Lorentz scalar, this could
enable the diffractive scattering to exhibit spin dependence at high
energies and undermine the Regge (nondiffractive) folklore that
$g_1(x \rightarrow 0) \sim$ const. as
has been commonly assumed in the experimental analyses.
\vskip 1.0cm
{\large{\bf $g_1(x)$ in the DLA of QCD}}
\vskip 0.5cm
In the DLA, the leading $\log \frac{1}{x}$ behaviour of $F_2(x)$ is driven
by the leading behaviour of the gluon-gluon splitting function at small $z$,
$P_{g g}(z) = 2/z$ and leads to the well-known result
$F_2(x \rightarrow 0) \sim \exp (k \sqrt{\ln\frac{1}{x}})$.

The helicity structure of the three-gluon vertex leads to a similar behaviour
for $\Delta g$, $\Delta q$ driven by $\Delta P_{g g} =4$
and hence $g_1$ (if we neglect complications from the anomaly term).
This yields $g_1 \sim \exp (\sqrt{2}k \sqrt{\ln\frac{1}{x}})$
and hence the relation
\begin{equation}
g_1 \sim [F_2]^{\sqrt{2}}
\label{dla}
\end{equation}
 The precise behaviour will
depend on the input polarised gluon distribution $\Delta G(x)$
which, in general, is expected to be non-zero \cite{cs}.
This provides an example of a naturally generated growth for $g_1$ at small
$x$ in QCD.
\vskip 1.0cm
\noindent{\large{\bf Empirical situation }}
\vskip 0.5cm
In $F_2(x,Q^2)$ the diffractive behaviour becomes dominant when
$x \lapproxeq 0.1$.
It is reasonable to assume that this is true also for $g_1(x)$; certainly
for $x \geq 0.1$ the valence quark model gives good predictions for $A_1(x)$
\cite{fecrgr2}
and there is no compelling reason to suspect that the valence - sea
transition occurs at radically different kinematic regions in the
different helicity states.

Now let us turn to the problem of using the assumed small $x$ behaviour
of $g_1^p(x)$ to extract a value for the integral $I_p(0,1) = \int_0^1
dx g_1^p(x)$ at $Q^2 = 10$ GeV$^2$ from the data. In Fig.1 the values
extracted from the asymmetry measurements by SMC \cite{smc} and EMC \cite{emc2}
are shown. These values assume that $A_1(x,Q^2)$ is independent of $Q^2$
and take recent fits \cite{mrsetc} for $F_2$ and $R$ to extract $g_1^p$
according to eq(\ref{g1defn}).
When the data on $A_1$ become more precise the proper analysis should
include the small $Q^2$ dependence expected from the evolution
equations\cite{anr}.
The new SMC data on the asymmetry $A_1^p$
continue to support the
predictions of valence quark models (VQM) for `large' $x$ and we can use these
to estimate the integral $I_p(0.135,1)$ reliably. The VQM curves in fig.1
give $I_p(0.135,1) = 0.080 \pm 0.008$.

To get an estimate of the low $x$ integral, we consider various possibilities
including those discussed above. The naive assumption $g_1(x) \sim $ constant
is not supported by the low $x$ SMC data, but the best fit of this type,
$g_1^p(x) = (0.35\pm 0.05)$,$ (x < 0.135)$ (see fig.1)
leads to $I_p(0,1)= 0.127\pm 0.010$,
$(\Delta q = 15\pm9\%$, if no higher twist present), to \cal O($\alpha_s$).
Next we consider three
examples where $xg_1(x)$ rises logarithmically
as $x \rightarrow 0$. For the $\log x$ behaviour given by
eq(\ref{g1logx}) the fit $xg_1^p(x) = (-0.14\pm 0.02)\ln x$,$(x < 0.135)$ leads
to $I_p(0,1)= 0.137\pm 0.011$,$(\Delta q = 24\pm10\%$). The two-gluon Pomeron
prediction \cite{basslan} of eq(\ref{g1bl}) gives a good fit to the low $x$
data with a coefficient $-0.085\pm$0.01 which is close to the preferred value
of $-0.09$. This gives $I_p(0,1)= 0.138\pm 0.011$,$(\Delta q = 25\pm11\%$).

Finally we consider an extreme point of view where a rapid rise at small $x$
is expected. General theorems on the high energy behaviour of the spin
dependent total cross sections show that {\it if} negative
signature cuts reach $J=1$ at $t=0$ there can be a leading contribution
to $xg_1(x) \sim 1/\log^2 x$ \cite{mt,kuti,kuti1}. Such a
behaviour was  discussed in an analysis of the first EMC results
\cite{fecrgr1}.
  Allowing such a rapid rise has been criticised \cite{leader}
but there seems to be no compelling argument for the decoupling of such
non-factorisable contributions to the amplitude.
Isoscalar $t-$channel exchanges with axial-vector quantum numbers,
as listed in eqs(4.1,4.2) of ref.\cite{heimann}, do include the possible
contributions from the negative signature cuts of refs\cite{mt,kuti,kuti1}.
 We are unaware of any
general theorems based on symmetry principles, angular momentum
etc. that forbid the above behaviour although it
may be that the magnitude of such contributions is indeed small or even
vanishing in specific dynamic models.

Phenomenologically
it is worth noting that the SMC data may be even more severe than
the $1/x\log^2 x$
 behaviour $-$ see fig. 1. Our analysis \cite{fecrgr1} of the initial
EMC data suggested the small $x$ region was consistent with
$xg_1^p(x) = 0.135/\ln^2 x$. The combined SMC and EMC data
prefer a parametrisation
$xg_1^p(x) = (0.17\pm 0.03)/\ln^2 x
,(x < 0.135)$ which leads to $I_p(0,1)= 0.165\pm 0.010$
 $ (\Delta q = 50\pm16\%)$.

Given the debatable nature of $g_1(x)$ as $ x \rightarrow 0$,
one could attempt to estimate the
integral $I_p(0,1)$ by simply fitting the small $x$ data to an arbitrary
power law plus a conventional constant term in order to assess
the range of uncertainty.
Even then the answer depends critically on the range of $x$
over which the fit is performed. For example taking $x < 0.135$ again,
the fit gives $g_1 \sim $const$ +x^{-2}$ which leads to
a divergent value for $I_p(0,1)$.

In any event this range of possibilities serves

(i) to illustrate that our limited understanding of the small $x$
region does allow for an estimate of the integral of $g_1^p(x)$ which is
entirely consistent with the original Ellis-Jaffe sum rule \cite{ej}
whose value, including $O(\alpha_s^2)$ corrections, is 0.172$\pm$0.009
at $Q^2 =$ 10 GeV$^2$.

(ii) as a challenge for future experiments to eliminate.

The resulting values
inferred for $\Delta q$ vary considerably
and so highlight the importance of being able
to discriminate between, at least, a roughly constant or falling $a_1$ pole
(non-diffractive or Lorentz scalar diffraction) on the one hand and a
(logarithmic) growth on the other.

Possible routes for resolving these questions include the following.

(a) Currently planned experiments
 \cite{hermes} giving precision data for $0.01 \leq x \leq 0.1$ which indicate
a
clear trend over this range and which tightly constrain continuation to
the less precise data from SMC at smaller $x$.

(b) Reduction of the systematic and statistical uncertainties in the SMC
data for $x \lapproxeq 0.05$ to confirm the apparent rise.

(c) Measurement of the sea polarisation directly via semi-inclusive production
of fast $K^-$ and $\pi$ \cite{cm,strik}

(d) Theoretical understanding of the rise at small $x$ in
$F_2(x,Q^2)$ and possible linkage with the Donnachie Landshoff description
being extended to a unified description involving spin dependence.

(e) Precise data for the deuteron at small $x$ where, if diffraction dominates,
$g_1^d(x)$ would be positive. Present data are not accurate enough to rule
out this possibility.

(f) Measurememt of the energy dependence of polarised $p p$ and polarised
(real) photoproduction asymmetries.

If any or all of these imply that there is significant non-trivial spin
dependence and growth in the diffractive region at small $x$, then
this may stimulate investigation of the possibility of creating longitudinally
polarised proton beams at HERA. Polarised electron - proton interactions
at HERA could turn out to have significant physics interest.
\vskip 1.0cm
\noindent{\large{\bf Acknowledgements}}
\vskip 0.5cm
We are grateful to
Steve Bass and Peter Landshoff for making their analysis available to us
before publication. We are grateful for discussions and suggestions
involving Jeff Forshaw, Erwin Gabathuler, Robert Heimann, Elliot Leader,
Andre Martin, Guido Martinelli, Al Mueller and Roger Phillips.

\bigskip

\vskip 1.0cm
\noindent{\large\bf Figure Caption}

\begin{itemize}
\item[Fig.\ 1] $g_1^p(x)$ at $Q^2 = 10$ GeV$^2$. Data are from refs
\cite{emc2,smc}.
\end{itemize}

\end{document}